\documentclass[%
 reprint,
 superscriptaddress,
 amsmath,amssymb,
 aps,
 pra,
]{revtex4-1}

\usepackage{graphicx}% Include figure files
\usepackage{dcolumn}% Align table columns on decimal point
\usepackage{bm}% bold math
\usepackage{multirow}
\usepackage{siunitx}
\usepackage{color}
\begin{document}

%\preprint{APS/123-QED}

\title{Tunable quantum photonics platform based on fiber-cavity enhanced single photon emission from two-dimensional hBN}

\author{Stefan H\"au\ss{}ler}
\thanks{These authors contributed equally to this work.}
\affiliation{Institute for Quantum Optics, Ulm University, Albert-Einstein-Allee 11, D-89081 Ulm, Germany}
\affiliation{Center for Integrated Quantum Science and Technology (IQst), Ulm University, Albert-Einstein-Allee 11, D-89081 Ulm, Germany}

\author{Gregor Bayer}
\thanks{These authors contributed equally to this work.}
\affiliation{Institute for Quantum Optics, Ulm University, Albert-Einstein-Allee 11, D-89081 Ulm, Germany}

\author{Richard Waltrich}
\affiliation{Institute for Quantum Optics, Ulm University, Albert-Einstein-Allee 11, D-89081 Ulm, Germany}

\author{Noah Mendelson}
\affiliation{School of Mathematical and Physical Sciences, University of Technology Sydney, \\
Ultimo, New South Wales 2007, Australia}

\author{Chi Li}
\affiliation{School of Mathematical and Physical Sciences, University of Technology Sydney, \\
Ultimo, New South Wales 2007, Australia}

\author{David Hunger}
\affiliation{Physikalisches Institut, Karlsruhe Institute of Technology, Wolfgang-Gaede-Stra\ss{}e 1, D-76131 Karlsruhe, Germany}

\author{Igor Aharonovich}
\affiliation{School of Mathematical and Physical Sciences, University of Technology Sydney, \\
Ultimo, New South Wales 2007, Australia}
\affiliation{ARC Centre of Excellence for Transformative Meta-Optical
Systems, Faculty of Science, University of Technology Sydney, Ultimo, New South Wales 2007, Australia}

\author{Alexander Kubanek}
\email[E-Mail: ]{alexander.kubanek@uni-ulm.de}
\affiliation{Institute for Quantum Optics, Ulm University, Albert-Einstein-Allee 11, D-89081 Ulm, Germany}
\affiliation{Center for Integrated Quantum Science and Technology (IQst), Ulm University, Albert-Einstein-Allee 11, D-89081 Ulm, Germany}

\date{\today}% It is always \today, today,
             % but any date may be explicitly specified

\begin{abstract}

Realization of quantum photonic devices requires coupling single quantum emitters to the mode of optical resonators. In this work we present a hybrid system consisting of defect centers in few-layer hBN grown by chemical vapor deposition and a fiber-based Fabry-Perot cavity. The sub $\SI{10}{\nano \meter}$ thickness of hBN and its smooth surface enables efficient integration into the cavity mode. We operate our hybrid platform over a broad spectral range larger than $\SI{30}{\nano \meter}$ and use its tuneability to explore different coupling regimes. Consequently, we achieve very large cavity-assisted signal enhancement up to $50$-fold and equally strong linewidth narrowing owing to cavity funneling, both records for hBN-cavity systems. Additionally, we implement an excitation and readout scheme for resonant excitation that allows us to establish cavity-assisted PLE spectroscopy. Our work marks an important milestone for the deployment of 2D materials coupled to fiber-based cavities in practical quantum technologies.

\end{abstract}

%\keywords{Suggested keywords}% Use showkeys class option if keyword
                              % display desired
\maketitle

%\tableofcontents

\section{Introduction}

Atomically thin van der Waals materials have recently attracted increasing attention as nanophotonic platforms due to their unique photophysical properties \cite{liu20192dmaterials}, lack of dangling bonds ideal for integration, and unparalleled potential for exploration of light-matter interactions at the nanoscale \cite{rivera2018interlayer, shimazaki2020strongly, chakraborty2019advances, caldwell2019photonics}. For example, the growing family of transition metal di-chalcogenides (TMDCs) is often used to study special exciton effects – e.g. Moire patterns \cite{blasius2020long} or exciton-polariton condensates \cite{dufferwiel2015exciton}. On the other hand, hexagonal boron nitride (hBN) has been explored \cite{watanabe2009farultraviolet, cassabois2016hexagonal} due to its ability to host bright and narrowband optically active luminescent defects that can act as single photon emitters (SPEs) %\cite{tran2015quantum} 
\cite{tran2016robust, bommer2019new} with very high quantum efficiency approaching $87 \, \%$ \cite{nikolay2019direct}. Recent progress has further accelerated interest in hBN as a quantum photonics platform. Outstanding milestones include demonstration of optically detected magnetic resonance \cite{gottscholl2020initialization, chejanovsky2019single}% \cite{mendelson2020identifying}
, a strong response to electrical and strain fields \cite{nikolay2019very, xia2019roomtemperature, mendelson2020strain}, photoluminescence upconversion via anti-Stokes process \cite{wang2018photoluminescence}, demonstration of Rabi-oscillations and resonance fluorescence \cite{konthasinghe2019rabi} as well as demonstrations of wafer scale growth of single crystal hBN \cite{chen2020wafer, lee2018wafer}. \\
One of the outstanding goals in the framework of SPEs in atomically thin materials is their efficient coupling to optical cavities \cite{flatten2018microcavity, lundt2016room} and photonics platforms \cite{kim2020hybrid}. Coupled emitter-cavity systems afford increased emission rates and enhanced collection efficiency, both representing critical goals for employing SPEs in scalable nanophotonic devices. Initial experiments with hBN have demonstrated the coupling of defect centers to dielectric photonic crystal cavities and microdisk resonators \cite{kim2018photonic, froech2019coupling, proscia2019scalable, shandilya2019hexagonal}, and cavity-enhanced single-photon generation with Fabry-Perot (FP) cavities in a compact architecture \cite{vogl2019compact}. However, the successful coupling yield of SPEs to the resonators with these geometries is still low. A critical issue has remained the wide range of emission energies observed for hBN SPEs, with zero-phonon lines (ZPL) known to range from $\sim 550-800 \, \mathrm{nm}$ \cite{jungwirth2017optical, tran2016robust, dietrich2018observation}, making efficient coupling to optical cavities with a fixed resonance wavelength challenging \cite{froech2019coupling, proscia2020microcavity}. \\
In this work we demonstrate a novel approach to overcome this issue by coupling hBN SPEs to a fiber based, FP open cavity. Our cavities exhibit a high finesse and small modal volume which is ideal for coupling to hBN SPEs. Most importantly, these cavities offer a unique advantage for the integration of 2D materials. Since the cavity is fully tunable over several $\SI{}{\micro \meter}$ and the focal area is large ($\sim \SI{1}{\micro \meter}$) enables the simultaneous study of many SPEs, with varying emission energies, within a single hBN layer. We utilize thin layers of hBN via chemical vapor deposition (CVD), which are advantageous over other hBN material sources as they provide a high surface quality to minimize scattering losses, and enable control over the photophysical properties of the defects during growth \cite{abidi2019selective, comtet2019widefield}.
\begin{figure*}[htbp]
	\includegraphics[width=0.98\textwidth]{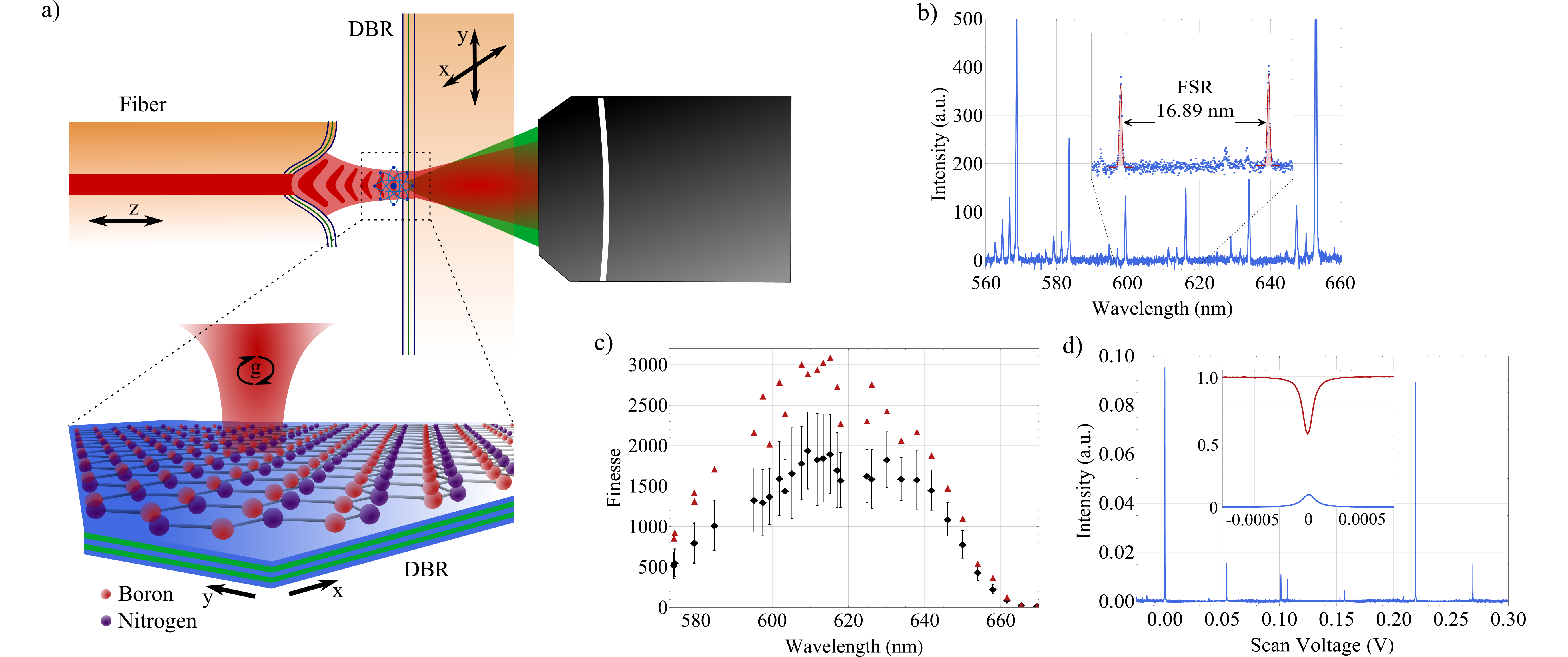}
		\caption{a) Design of the micro resonator setup. A macroscopic plane mirror and the tip of a glass fiber with concave structure form a plano-concave resonator. A CVD grown hBN layer is placed on the plane mirror forming a hybrid quantum emitter fiber-cavity system. b) Cavity transmission spectrum in the wavelength range from $\SI{560}{\nano \meter}$ to $\SI{660}{\nano \meter}$ revealing a cavity free spectral range (FSR) of $\Delta \lambda_{\text{FSR}} = \SI{16.89}{\nano \meter}$ at $\SI{600}{\nano \meter}$. c) Cavity finesse $\mathcal{F}$ as a function of the wavelength (black data) determined by scanning cavity measurements. Maximum values (red triangles) $> 3000$ are reached between $\SI{590}{\nano \meter}$ and $\SI{640}{\nano \meter}$, while the finesse decays to the higher (lower) wavelength side due to the limited mirror reflectivities. d) Cavity transmission versus piezo voltage over one cavity FSR. We observe strong coupling to the fundamental mode, while higher order modes are weakly pronounced. The inset shows the transmitted (blue) and reflected (red) signal at resonance. The maximum coupling efficiency visible in the reflection measurement is limited by the impedance mismatch of the fiber mirror and the cavity mode.}
	\label{fig:figure1}
\end{figure*}

\section{Experimental Platform}

Spectroscopic measurements were performed using a custom-built confocal microscope with an $\mathrm{NA} = 0.7$ microscope objective combined with a micro resonator setup consisting of a macroscopic plane mirror and a fiber mirror. Both, the plane and the fiber mirror are provided with a highly reflective dielectric coating to form a distributed Bragg reflector (DBR). The DBR stack of the plane mirror is designed for high reflectivities in the spectral range common for hBN SPEs (i.e. from $\SI{560}{\nano \meter}$ to $\SI{740}{\nano \meter}$) and has a high index termination, which leads to a field node on the surface of the mirror. The fiber mirror is fabricated with a concave structure having a radius of curvature ($\mathrm{ROC}$) of $R_{c} = \SI{35}{\micro \meter}$, which was produced by laser ablation from a high power $\text{CO}_{2}$ laser before coating. The ROC of the structure is estimated from a parabolic fit to an interferometer image enabling resonator lengths $L_{0}$ of more than $\SI{30}{\micro \meter}$ in stable operation. The homogeneous ablation of material from the high intensity infrared laser leads to a relatively smooth surface enabling high finesse values \cite{hunger2010afiber}. The reflectivity of the fiber mirror ranges from $\SI{560}{\nano \meter}$ to $\SI{660}{\nano \meter}$. Both dielectric coatings are transmissive for green ($\SI{532}{\nano \meter}$) laser light, enabling off-resonant excitation of color centers inside the resonator. The length of the fiber cavity (z-direction) is tunable via the fiber mirror side while the plane mirror can be adjusted in the lateral direction (x- and y-direction). A schematic representation of the platform is shown in figure \ref{fig:figure1}(a). Switching between scannable confocal microscope and fiber-cavity based collection mode is possible by flipping the plane mirror. First we characterize the bare fiber-cavity (without hBN) via the fiber mirror side. We measure the spectrum of the fiber resonator by recording the transmission of a broadband light source with a grating spectrometer ($\SI{1200}{groves/mm}$) for different cavity lengths. For an exemplary length of $L_{0} = \SI{10.66}{\micro \meter}$ the transmission spectrum is depicted in figure \ref{fig:figure1}(b) resulting in a cavity free spectral range (FSR) of $\Delta \lambda_{\mathrm{FSR}} = \SI{16.89}{\nano \meter}$ at $\SI{600}{\nano \meter}$ ($\Delta \nu_{\mathrm{FSR}} = \SI{14.07}{\tera \hertz}$). The spectral position of the fundamental modes for a given cavity length are utilized to calculate and compensate the piezo hysteresis by applying a third order polynomial function to the measured data (details see Supplemental Material). \\
We determine the cavity finesse $\mathcal{F}$ by measuring the transmission of a narrowband ($\SI{100}{\kilo \hertz}$) dye ring laser system while continuously scanning the cavity length. The cavity finesse is defined as the ratio of the FSR and the FWHM linewidth of the resonances $\mathcal{F} = \mathrm{FSR}/\mathrm{FWHM}$. High finesse values of approximately $3000$ are obtained in the wavelength range between $\SI{590}{\nano \meter}$ and $\SI{640}{\nano \meter}$, while the finesse decreases for higher (lower) wavelengths due to decreasing mirror reflectivity of the fiber (plane) mirror (figure \ref{fig:figure1}(c)). The cavity length can be varied within the theoretically predicted stability regime without significant changes of the finesse (see Supplemental Material). However in this work we focus on short cavity lengths ($< \SI{15}{\micro \meter}$) for increased coupling between SPE and cavity mode (for detailed calculations see Supplemental Material). The transmission through the scanning cavity (figure \ref{fig:figure1}(d)) displays pronounced fundamental modes, while higher order modes appear strongly suppressed. The maximum coupling efficiency is extrapolated from the reflection measurement (red curve in the inset of figure \ref{fig:figure1}(d)) and yields $\si{45}{\%}$. The efficiency is limited by the impedance mismatch between the fiber end and the cavity mode.

\section{Sample characterization}

\begin{figure}[htbp]
	\includegraphics[width=0.48\textwidth]{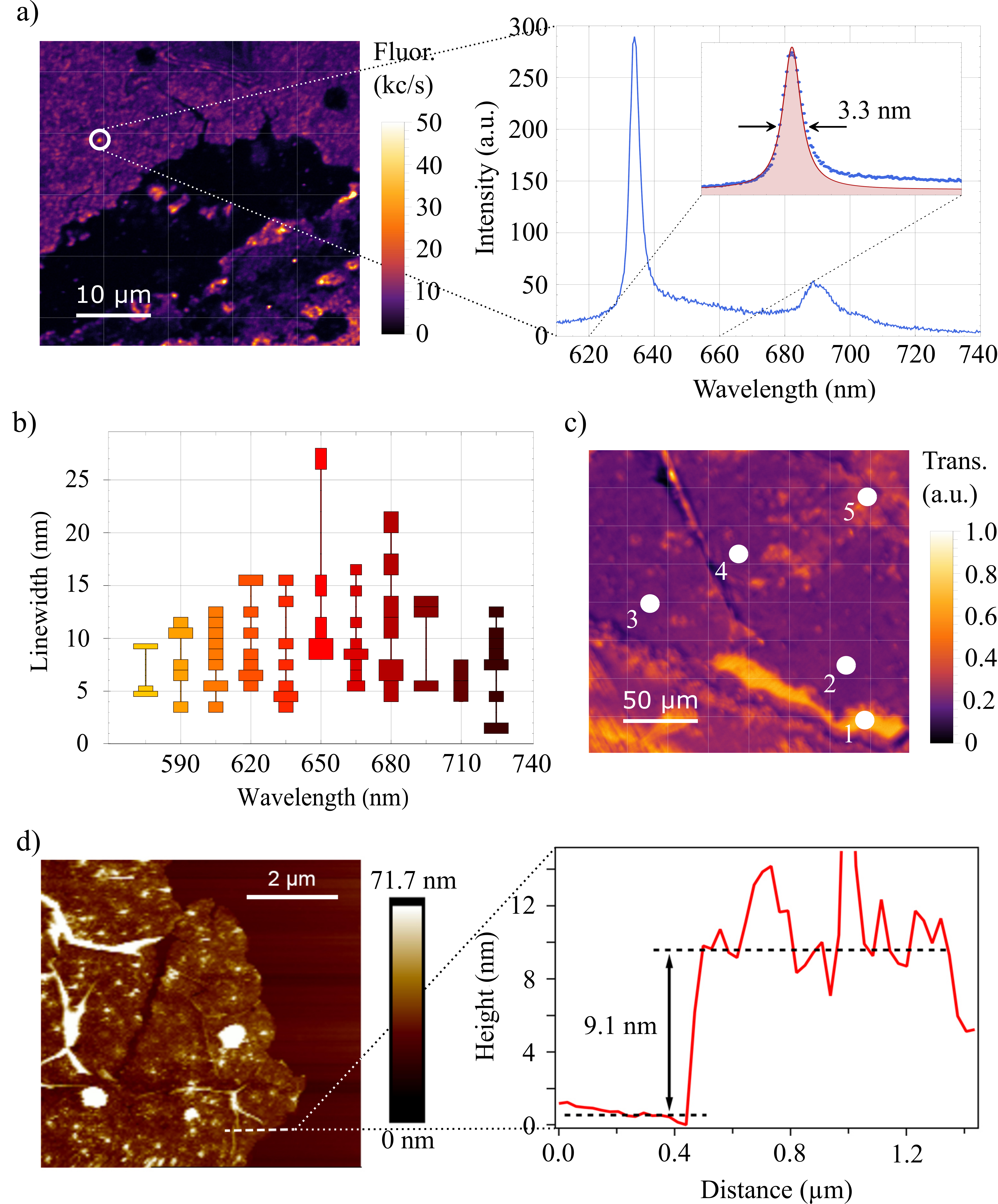}
		\caption{a) Confocal image of a $50 \times 50 \, \si{\micro \meter}^{2}$ region of the CVD grown hBN layer on the DBR. The PL spectrum of the marked defect exhibits strong ZPL emission at $\lambda \approx \SI{632}{\nano \meter}$ with a FWHM linewidth of $\delta \lambda = \SI{3.3}{\nano \meter}$ and phonon sideband (PSB) up to $\approx \SI{740}{\nano \meter}$. b) Spectral distribution of defects in the CVD grown hBN layer ranges from $\SI{570}{\nano \meter}$ to $\SI{740}{\nano \meter}$. The FWHM linewidths spread from $\SI{3}{\nano \meter}$ to $\SI{27}{\nano \meter}$. In this statistic the majority of defects in the range from $\SI{570}{\nano \meter}$ to $\SI{590}{\nano \meter}$ is excluded for clearance. c) Cavity transmission map of a broadband light source to investigate the performance of the hBN layer in the cavity. The cavity finesse is measured on the marked spots to estimate the scattering behavior of the layer. d) Atomic force microscopy scan of a $10 \times 10 \, \si{\micro \meter}^{2}$ region of the CVD grown hBN layer. The height trace taken from the white dashed line yields a layer thickness of $t_{\mathrm{hBN}} \approx \SI{9.1}{\nano \meter}$.}
	\label{fig:figure2}
\end{figure}

The sample is fabricated using large scale thin layers of hBN grown by CVD on copper, as described in detail in reference \cite{mendelson2019engineering}. The hBN layer is sandwiched between two PMMA layers of $\SI{95}{\nano \meter}$ thickness to position the defects in the field maximum of the cavity mode. The hBN layer is first examined via confocal microscopy at room temperature using an off-resonant laser (DPSS) at $\SI{532}{\nano \meter}$. An exemplary confocal scan of a $50 \times 50 \, \si{\micro \meter}^{2}$ region is shown in figure \ref{fig:figure2}(a). 
We record photoluminescence (PL) spectra of numerous defects revealing that the hBN hosts a large variety of defects covering a broad wavelength range from $\approx \SI{570}{\nano \meter}$ up to $\SI{740}{\nano \meter}$. The majority of defects emit between $\SI{570}{\nano \meter}$ and $\SI{590}{\nano \meter}$, as reported in the references \cite{mendelson2019engineering,abidi2019selective, comtet2019widefield, stern2019spectrally}. A PL spectrum of such a defect (marked spot in figure \ref{fig:figure2}(a)) features pronounced zero-phonon line (ZPL) emission centered at $\lambda \approx \SI{632}{\nano \meter}$ with a FWHM linewidth of $\delta \lambda = \SI{3.3}{\nano \meter}$. The linewidth of the investigated defects spreads from $\SI{3}{\nano \meter}$ to more than $\SI{20}{\nano \meter}$ (figure \ref{fig:figure2}(b)). \\
The scattering losses due to the inserted hBN layer inside the resonator are investigated by recording a transmission map of the broadband light source (see figure \ref{fig:figure2}(c)). Additionally we measure the finesse of the cavity on several regions on the hBN layer (marked spots in figure \ref{fig:figure2}(c)) and extract the scattering losses introduced by the layer, as it has been done for a diamond membrane cavity system \cite{haeussler2019diamond}, resulting in:
\begin{center}
\begin{tabular}{|c|c|c|}
\hline
\rule{0pt}{10pt} \ \ \ Spot number \ \ \ & \ \ \ Finesse \ \ \ & \ \ \ Scattering Losses \ \ \ \\
\hline
\hline
1 (mirror) & 2248 & 1400 ppm \\
2 & 465 & 5360 ppm \\
3 & 375 & 6980 ppm \\
4 & 195 & 14710 ppm \\
5 & 886 & 2150 ppm \\
\hline
\end{tabular}
\end{center}
This yields an average rms surface roughness of $\sigma = \SI{4.1 \pm 1.1}{\nano \meter}$ of the hBN layer. Furthermore atomic force microscopy (AFM) is applied - measured on SiO$_{2}$ prior to hBN encapsulation between the two PMMA layers - to determine the overall height of the hBN layer. An AFM scan of the hBN layer edge ($10 \times 10 \, \si{\micro \meter}^{2}$) with a resolution of $512 \times 512 \, \si{px}$ is shown in figure \ref{fig:figure2}(d). The height trace along the white dashed line yields an average layer thickness of $t_{\mathrm{hBN}} \approx \SI{9.1}{\nano \meter}$ and a surface roughness of few $\SI{}{\nano \meter}$, in agreement with the extracted scattering losses. We thus conclude that the surface quality of the hBN layer is suitable for an integration into the cavity mode. We note that variations in surface roughness, primarily a result of wrinkling during the transfer process, lead to variations in the effective cavity finesse, with smoother areas providing higher finesse values. However, additional optimization of the growth process, or additional polishing of the hBN layer after growth can further improve the obtained results.

\section{Coupled system}

\begin{figure*}[htbp]
	\includegraphics[width=0.98\textwidth]{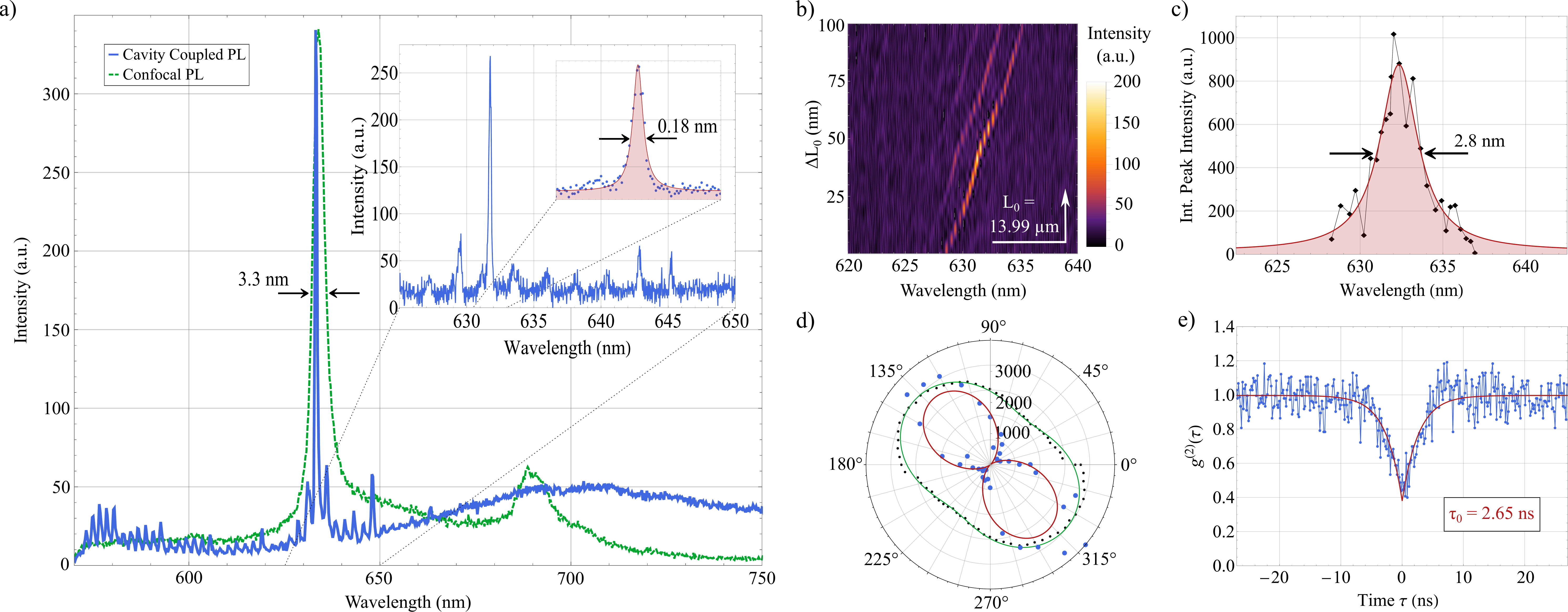}
		\caption{a) Cavity coupled (blue) and free space (green) spectrum of the $\lambda = \SI{632}{\nano \meter}$ defect (PL spectrum in figure \ref{fig:figure2}(a)) represents pronounced funneling of the emission into a single optical mode of the cavity with a more than $18$-fold reduced ZPL linewidth. The inset shows the high resolution image from where the linewidth of $\delta \lambda = \SI{0.18}{\nano \meter}$ is extracted. b) Cavity coupled fluorescence spectra of the $\lambda = \SI{632}{\nano \meter}$ defect (figure \ref{fig:figure2}(a)) for different cavity lengths $L_{0} + \Delta L_{0}$ starting at $L_{0} = \SI{13.99}{\micro \meter}$. c) Integrated cavity fluorescence intensity of the $\lambda = \SI{632}{\nano \meter}$ defect for different cavity lengths. Peak intensity is observed at the center of the defect ZPL. d) Emission polarization of the $\lambda = \SI{632}{\nano \meter}$ defect (blue data, red fit) shows preferential emission into the polarization axis of the cavity (black data, green fit) with high polarization contrast. e) Second-order autocorrelation measurement of the $\lambda = \SI{632}{\nano \meter}$ defect clearly indicates single photon emission.}
	\label{fig:figure3}
\end{figure*}

The plane mirror is now moved laterally to overlap the position of individual defect centers in the hBN layer with the cavity mode. Off-resonant excitation (at $\SI{532}{\nano \meter}$) as well as the collection of the cavity mode is done via the plane mirror side. We observe coupling of the defect presented in figure \ref{fig:figure2}(a) with a ZPL centered at $\lambda = \SI{632}{\nano \meter}$ to a single mode (TEM$_{00}$) of the fiber-cavity. The cavity-coupled hBN spectrum is presented in figure \ref{fig:figure3}(a) (low resolution ($\SI{0.52}{\nano \meter}$)), while the inset displays the corresponding high resolution ($\SI{0.06}{\nano \meter}$) spectrum. The background signal visible in figure \ref{fig:figure3}(a) originates from fiber fluorescence ($\SI{650}{\nano \meter}$ to $\SI{750}{\nano \meter}$) and surrounding defects ($\SI{570}{\nano \meter}$ to $\SI{600}{\nano \meter}$). Cavity filtering can be observed by the more than $18$-fold reduction of the ZPL linewidth of the defect now dictated by the corresponding cavity linewidth (see inset in figure \ref{fig:figure3}(a)). Note that all broadening effects like the cavity jitter during the acquisition time are included. We further tune the cavity resonance (by scanning the cavity length) over the ZPL frequency of the defect and observe a resonance at $\lambda \approx \SI{632}{\nano \meter}$ (figure \ref{fig:figure3}(b)). Integrating over all spectra for the scan in figure \ref{fig:figure3}(b) yields again approximately the ZPL linewidth of the defect in free-space emission (figure \ref{fig:figure3}(c)). \\
Furthermore, we observe linear polarized emission of the coupled defect with high polarization contrast aligned with the polarization axis of the cavity (see figure \ref{fig:figure3}(d)) leading to increased coupling between defect and cavity mode. The high polarization contrast mainly originates from the defect itself, since in the regime with $\mathcal{F} < 3000$ the cavity polarization modes are degenerate, lowering the polarization contrast of the cavity. A second-order autocorrelation measurement at low excitation power far from saturation yields single photon emission from the coupled system with $g^{(2)}(0) = 0.4$ and an extracted lifetime of $\tau_{0} = \SI{2.65}{\nano \second}$ (figure \ref{fig:figure3}(e)). \\

\begin{figure*}[htbp]
	\includegraphics[width=0.98\textwidth]{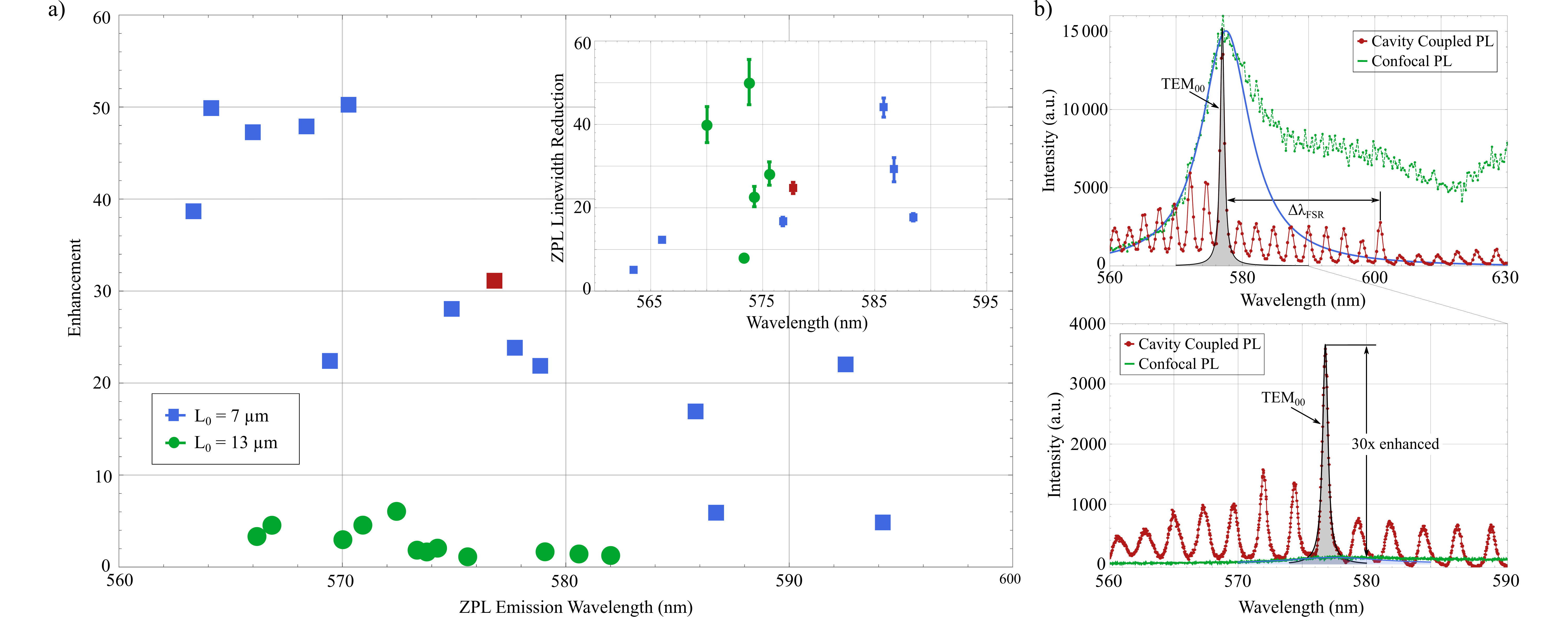}
		\caption{a) Spectral enhancement for different fluorescing defects in the wavelength range from $\SI{560}{\nano \meter}$ to $\SI{600}{\nano \meter}$ for two different cavity lengths $L_{0} = \SI{7}{\micro \meter}$ (blue data) and $L_{0} = \SI{13}{\micro \meter}$ (green data). The data of the defect in (b) is marked red. The inset shows the corresponding reduction of the ZPL linewidth for selected defects. b) Cavity coupled spectrum (red) and free space PL spectrum (green) of a defect with its ZPL centered at $\lambda \approx \SI{577}{\nano \meter}$ (top). The free space emission is scaled to the intensity of the cavity coupled emission for better visibility. The fluorescence of the ZPL in the cavity is spectrally enhanced by a factor of $30$ compared to the free space emission as it can be extracted from the high resolution image (bottom).}
	\label{fig:figure4}
\end{figure*}

To further establish cavity-assisted enhancement and cavity funneling we investigate the coupling of numerous defects at different wavelengths to the cavity mode. In each case the TEM$_{00}$-mode of the cavity is tuned into resonance with the ZPL of the particular defect. As a result we observe strong cavity funneling of the ZPL emission of these defects into the cavity mode. To classify the effect of the cavity we extract the spectral enhancement of each ZPL in the cavity mode as well as the reduction of its linewidth by comparing the free space emission with the cavity coupled emission. In total spectral enhancement is studied for more than $20$ defects in the wavelength range between $\SI{560}{\nano \meter}$ and $\SI{600}{\nano \meter}$ for two different cavity lengths $L_{0} = \SI{7}{\micro \meter}$ and $L_{0} = \SI{13}{\micro \meter}$ (see figure \ref{fig:figure4}(a)). 
As expected the largest enhancement values are obtained for very short cavity lengths. At $L_{0} = \SI{7}{\micro \meter}$ most defects are more than $20$-fold enhanced, while selected defects exhibit enhancement values of up to $50$. At $L_{0} = \SI{13}{\micro \meter}$ however the largest enhancement ($6.3$) is observed for a defect centered at $\approx \SI{572}{\nano \meter}$. \\
The large spectral enhancement in a well-defined optical mode of the cavity is based on two effects. First, the Purcell factor explains the increased emitter-cavity coupling growing with shorter cavity length. We experimentally confirm this by measuring the spectral enhancement for the two cavity lengths $L_{0} = \SI{7}{\micro \meter}$ and $L_{0} = \SI{13}{\micro \meter}$. We estimate the effective Purcell factor in the bad emitter regime, c.f. reference \cite{grange2015cavity}, for a strongly dissipative quantum emitter coupled to a high quality cavity for different cavity lengths resulting in a Purcell factor of $F_{p} \approx 1$ for  $L_{0} = \SI{7}{\micro \meter}$, $2.3$ times higher than for $L_{0} = \SI{13}{\micro \meter}$ (details see Supplemental Material). Second, the slightly asymmetric coating of the two cavity mirrors ($T_{\mathrm{plane}} > T_{\mathrm{fiber}}$ for wavelengths $< \SI{600}{\nano \meter}$) and the overall higher transmission at lower wavelength is designed for high out-coupling rates ($T_{\mathrm{plane}}/(T_{\mathrm{plane}} + T_{\mathrm{fiber}} + L)$) on the planar mirror side (c.f. also figures \ref{fig:figure1}(b) and (c)). \\
In both cases (free space and cavity coupled) we use identical excitation conditions, in particular the same intensity and numerical aperture optics. Cavity effects of the excitation light can be neglected due to high mirror transmission for the green $\SI{532}{\nano \meter}$ excitation laser. To illustrate the determination of the enhancement factor the cavity coupled spectrum of a defect at $\lambda \approx \SI{577}{\nano \meter}$ (red data in figure \ref{fig:figure4}(a)) is shown in figure \ref{fig:figure4}(b). Multiple cavity modes are fed by the broadband fluorescence of the defect center. The strongest mode overlaps with the central wavelength of the ZPL at $\lambda = \SI{577}{\nano \meter}$, demonstrating a spectral enhancement factor of $30$ compared to the free space emission (green data in figure \ref{fig:figure4}(b)). \\

\begin{figure}[htbp]
	\includegraphics[width=0.48\textwidth]{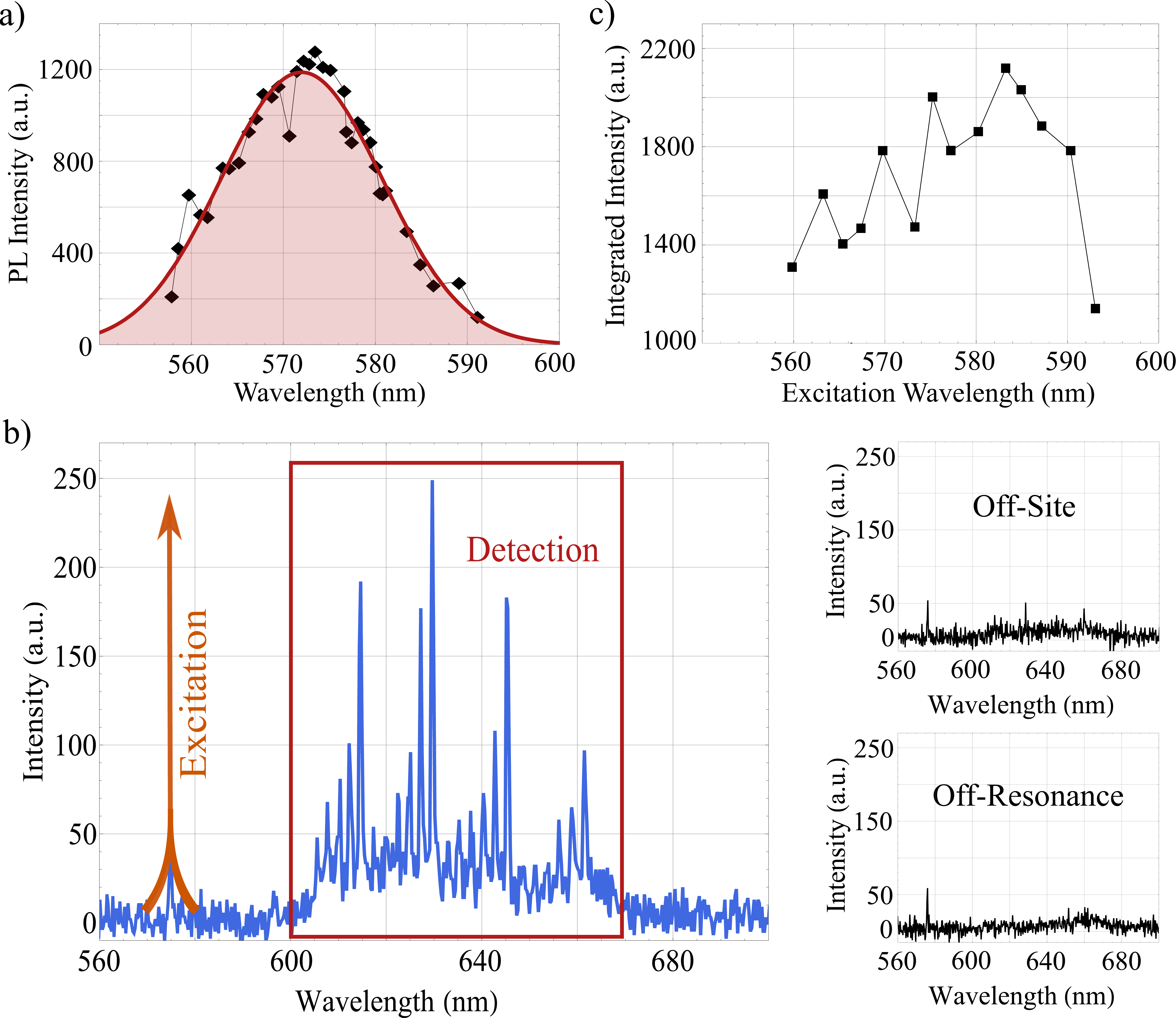}
		\caption{a) Integrated cavity coupled PL intensity of the ZPL under off-resonant excitation for different cavity lengths. b) Resonant excitation scheme, which shows the excitation at approx. the center of the ZPL at $\lambda = \SI{575}{\nano \meter}$ and the detection of the side band in the range $\SI{600}{\nano \meter}$ to $\SI{680}{\nano \meter}$. We do not observe any signal from the defect with the excitation laser on-resonance with the cavity but the cavity mode laterally detuned from the defect (bottom left) nor with the cavity mode on-site but the laser off-resonance (bottom right). c) Excitation efficiency of the defect over the excitation wavelength. We observe a strong decay starting at $\approx \SI{590}{\nano \meter}$. On the lower wavelength side the decay is not as steep due to the higher probability for off-resonant excitation.}
	\label{fig:figure5}
\end{figure}

Recently mechanically isolated defect centers have been demonstrated under resonant excitation yielding narrow linewidths up to $\SI{60}{\mega \hertz}$ at room temperature \cite{dietrich2020solidstate, hoese2020mechanical}. To access this regime we establish a method to perform cavity-assisted PLE spectroscopy comparable to the scheme presented in reference \cite{casabone2018cavity}). We resonantly excite a defect with its ZPL centered at $\lambda \approx \SI{575}{\nano \meter}$ (c.f. integrated cavity coupled PL spectrum under off-resonant excitation in figure \ref{fig:figure5}(a)) and detect the PSB in the lower order fundamental $q$ modes of the cavity as shown in figure \ref{fig:figure5}(b). 
As an excitation light source we use here a tunable optical-parametric oscillator (OPO) with second-harmonic-generation (SHG) unit that is tuned on resonance with the cavity. To our knowledge this is the first evidence of cavity-assisted PLE for defects in hBN. To prove the resonant excitation, we detune the excitation laser from the cavity resonance observing no signal. In addition, we do not observe emission when the cavity is on resonance with the excitation laser but laterally displaced from the position of the defect (by shifting the plane mirror). We further measure the inhomogeneous linewidth of the defect center under resonant and near-resonant excitation by tuning the excitation wavelength over the ZPL (figure \ref{fig:figure5}(c)). The displayed integrated intensity hereby corresponds to the cumulative intensity of the fundamental modes in the PSB (c.f. figure \ref{fig:figure5}(b)). The inhomogeneous linewidth observed in PLE spectroscopy is in agreement with the PL linewidth. However, on the short-wavelength side of the ZPL the near-resonant excitation via acoustic phonon modes becomes apparent. \\
In the presented case, the electron-phonon density of states for the acoustic phonon branch is gap-less and merges into the ZPL. This leads to a reduced excitation efficiency under resonant excitation at room temperature. At the same time, the emission enhancement induced by the cavity is limited by the strong dephasing of the investigated defect centers. We extrapolate the achieved cavity enhancement for a defect exhibiting Fourier-Transform limited lines at room temperature as it is reported in the references \cite{dietrich2020solidstate, hoese2020mechanical} resulting in Purcell factors $> 100$ assuming ideal emitter-cavity coupling (details see Supplemental Material).

\section{Conclusion}

In summary, we have studied the coupling of quantum emitters in few-layer CVD grown hBN to FP open cavities. We have demonstrated record cavity-enhanced emission for individual defect centers in hBN up to $50$-fold enhancement and up to $50$-fold linewidth reduction due to cavity funneling. Importantly, our hybrid system offers a broad spectral tuning range ideal for coupling to the wide range of ZPL emission energies common to hBN, demonstrated here by coupling single defect center with emission wavelength ranging from $\SI{565}{\nano \meter}$ to $\SI{635}{\nano \meter}$. Our work paves the way for indistinguishable single photon emission at high rates with potential for enhancing spin readout and for fiber-based quantum photonics with applications in quantum communication. \\
We further establish cavity-assisted PLE spectroscopy which can be utilized in the future to characterize mechanically isolated defect centers in hBN at room temperature \cite{dietrich2020solidstate} with critically enhanced photon flux. Making use of cavity-assisted excitation is essential to reduce the necessary excitation power. Since spectral diffusion is power-dependent, our platform can be utilized to overcome spectral instability, which is a prevalent issue in the deployment of hBN SPEs for practical applications. Dissipation in our coupled system is currently limited by rapid emitter dephasing. Extending our work to mechanically isolated defect centers could eliminate emitter dephasing reaching the Fourier-Transform limit and resulting in emitter linewidth of $\gamma_{0} \approx \SI{60}{\mega \hertz}$ at room temperature. Our results therefore hold the unique capability to establish a strongly coupled solid-state system with a cooperativity exceeding $C_{0} > 100$ at room temperature.

\section*{Acknowledgements}

Experiments performed for this work were operated using the Qudi software suite \cite{binder2017qudi}. SH and AK acknowledge support of IQst. AK acknowledges support from the Baden-W\"urttemberg Stiftung in project Internationale Spitzenforschung. IA acknowledges the generous support provided by the Alexander von Humboldt Foundation and the Australian Research council (DP180100077). DH acknowledges support by the Karlsruhe School of Optics \& Photonics (KSOP).

\section*{Author Contributions}

SH, GB, RW and AK conceived the experiments, performed the measurements and evaluated the data. DH fabricated the fiber mirrors. NM, CL and IA prepared the hBN samples and performed the transfer onto the mirrors. The manuscript was written by SH and AK and all authors discussed the results and contributed to the manuscript.

%\nocite{*}

\bibliography{Quantum_photonics_platform_based_on_FC_enhanced_single_photon_emission_from_2D_hBN}
% Produces the bibliography via BibTeX.

\end{document}